\documentclass[smallabstract,smallcaptions]{dccpaper}
\usepackage{graphicx}
\usepackage{amsmath}
\usepackage{amssymb}
\usepackage{color}
\usepackage{url}
\usepackage{titlesec}
\usepackage{subcaption}
\usepackage{svg}
\usepackage{caption}

\titleformat{\subsubsection}[runin]
{\normalfont\itshape}
{\thesubsubsection. }{.5em}{}[.]
\titlespacing{\subsubsection}
{\parindent}{0.5ex plus .1ex minus .2ex}{3pt}

\newlength{\figurewidth}
\newlength{\smallfigurewidth}

\AtBeginDocument{%
  \providecommand\BibTeX{{%
    \normalfont B\kern-0.5em{\scshape i\kern-0.25em b}\kern-0.8em\TeX}}}

\usepackage{todonotes}

\usepackage{amsmath, amsthm}

\usepackage{cases}
\usepackage{graphicx}
\usepackage{hyperref, url}
\usepackage{adjustbox}
\usepackage{booktabs, multirow}
\usepackage{caption}
\usepackage{subcaption}
\usepackage{enumitem}
\usepackage{lipsum}
\usepackage{setspace}
\usepackage[xcolor]{changebar}
\usepackage[normalem]{ulem}

\makeatletter
\newcommand{\cboff}{\let\end@dblfloat\ltx@end@dblfloat}
\newcommand{\cbon}{\let\end@dblfloat\cb@end@dblfloat}
\makeatother

\newcommand{\deleted}[1]{%
  \cbcolor{red}
  }%

\usepackage{thmtools}
\declaretheoremstyle[%
  spaceabove=4pt,%
  spacebelow=4pt,%
  headfont=\normalfont\itshape,%
  postheadspace=1em,%
  qed=\qedsymbol%
]{mystyle}

\usepackage{footnote}
\makesavenoteenv{tabular}
\makesavenoteenv{table}

\usepackage{xspace}

\usepackage{tikz}
\usetikzlibrary{calc}

\usepackage{array}
\usepackage{booktabs}
\usepackage{tabu}

\begin{document}

\title{
\large
\textbf{Accurate Performance Modeling And Uncertainty Analysis of Lossy Compression in Scientific Applications}
}


\author{%
Youyuan Liu$^{\ast}$, Taolue Yang$^{\ast}$, Sian Jin$^{\ast}$ \\[0.5em]
{\small\begin{minipage}{\linewidth}\begin{center}
\begin{tabular}{cc}
$^{\ast}$ Temple University\\
Philadelphia, PA, 19121, USA\\
\url{{youyuan.liu, taolue.yang, sian.jin}@temple.edu}
\end{tabular}
\end{center}\end{minipage}}
}

\newcommand{\tmp}{Temple University}
\newcommand{\iu}{Indiana University}
\newcommand{\anl}{Argonne National Laboratory}
\newcommand{\lanl}{Los Alamos National Laboratory}
\newcommand{\lbl}{Lawrence Berkeley National Laboratory}
\newcommand{\pnnl}{Pacific Northwest National Laboratory}
\newcommand{\fsu}{Florida State University}
\newcommand{\inria}{Inria \& LIP, ENS Lyon}
\newcommand{\ens}{Inria \& LIP, ENS Lyon}

\newcommand{\AFFIL}[4]{%
     \affiliation{\small
         \institution{#1}
         \city{#2}\state{#3}\country{#4}
     }
     }

\newcommand{\TMP}{\AFFIL{\tmp}{Philadelphia}{PA}{USA}}
\newcommand{\IU}{\AFFIL{\iu}{Bloomington}{IN}{USA}}
\newcommand{\ANL}{\AFFIL{\anl}{Lemont}{IL}{USA}}
\newcommand{\LANL}{\AFFIL{\lanl}{Los Alamos}{NM}{USA}}
\newcommand{\LBL}{\AFFIL{\lbl}{Berkeley}{CA}{USA}}
\newcommand{\PNNL}{\AFFIL{\pnnl}{Richland}{WA}{USA}}
\newcommand{\FSU}{\AFFIL{\fsu}{Tallahassee}{FL}{USA}}
\newcommand{\INRIA}{\AFFIL{\inria}{Lyon}{}{France}}
\newcommand{\ENS}{\AFFIL{\ens}{Lyon}{}{France}}









\maketitle
\thispagestyle{empty}




\begin{abstract}
Scientific applications typically generate large volumes of floating-point data, making lossy compression one of the most effective methods for data reduction, thereby lowering storage requirements and improving performance in large-scale applications.
However, variations in compression time can significantly impact overall performance improvement, due to inaccurate scheduling, workload imbalances, etc.
Existing approaches rely on empirical methods to predict the compression performance, which often lack interpretability and suffer from limitations in accuracy and generalizability. 
In this paper, we propose surrogate models for predicting the compression time of prediction-based lossy compression and provide a detailed analysis of the factors influencing time variability with uncertainty analysis. 
Our evaluation shows that our solution can accuratly predict the compression time with 5\% average error across six scientific datasets. It also provides accurate 95\% confidence interval, which is essential for time-sensitive scheduling and applications.
\end{abstract}

\section{Introduction}
Large-scale scientific applications have become a crucial component in modern scientific research, often producing vast amounts of data. For example, a single Nyx~\cite{nyx} cosmological simulation at a resolution of 4096 × 4096 × 4096 cells can generate over 4.2 PB data per simulation. 
These simulations are often repeated, further increasing the data volume and presenting two key challenges: (1) storing such massive datasets is difficult, even with supercomputing resources; (2) limited I/O bandwidth makes data transfer and processing time-consuming.
Lossy compression offers an effective solution by significantly reducing data size while maintaining data distortion within acceptable limits. 
This is especially true compared to lossless compression, which typically reduces continuous floating-point data by $\sim$2$\times$.
New generation of lossy compressors, such as SZ~\cite{di2016fast, tao2017significantly, sz18, sz3} and ZFP~\cite{zfp}, have been widely adopted in various systems and scientific applications to reduce data sizes and enhance performance~\cite{cappello2025multifacets}.

\begin{figure}[]
    \centering
    \begin{subfigure}{0.45\linewidth}
        \centering
        \includegraphics[width=\linewidth]{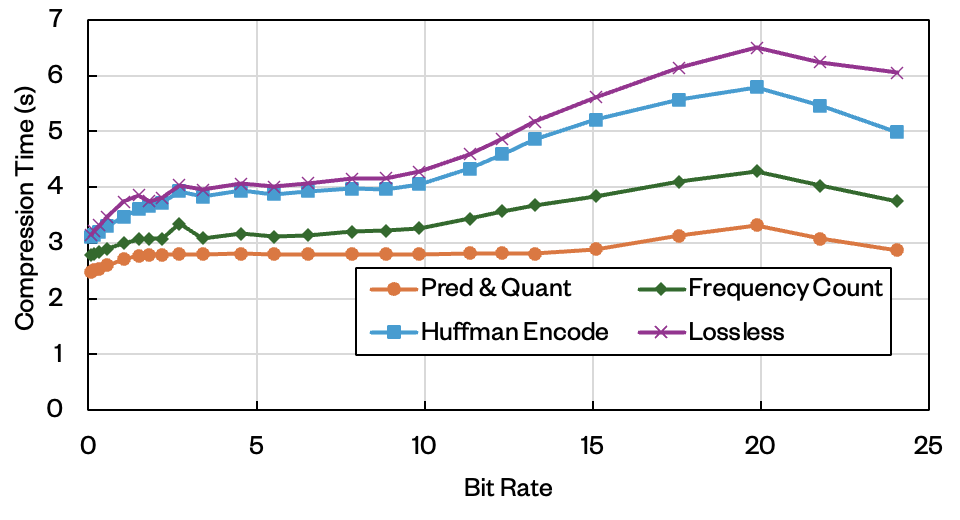}
        \caption{\scriptsize Different bitrate with Nyx's velocity\_x dataset.}
        \label{fig:stack}
    \end{subfigure}
    \begin{subfigure}{0.45\linewidth}
        \centering
        \includegraphics[width=\linewidth]{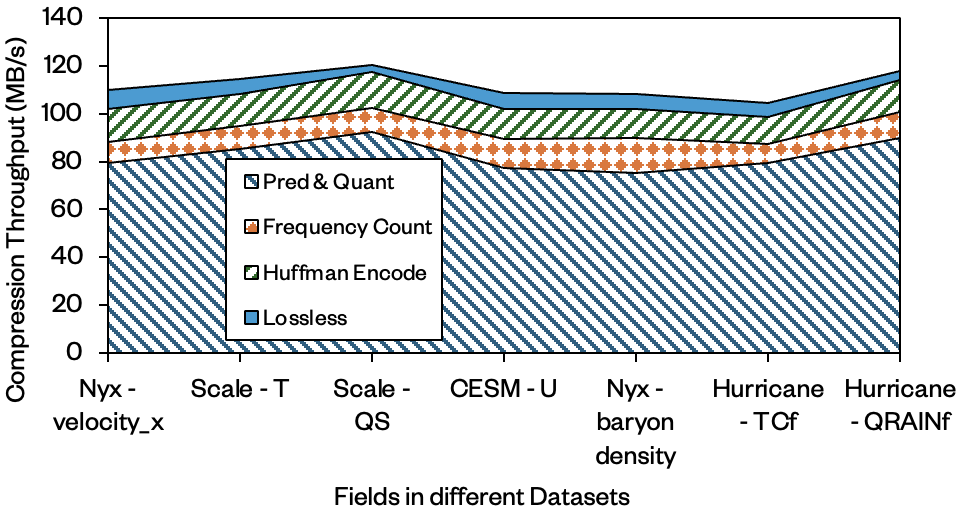}
        \caption{\scriptsize different scientific datasets.}
        \label{fig:different-dataset}
    \end{subfigure}
    \caption{Compression throughput varies significantly with different bitrate and datasets. Internally, the throughput is stacked by different stages based on their execution time.}
    \label{fig:comp-time}
\end{figure}

Compression throughput is often one of the most critical metrics, as it directly impacts execution time and plays a key role in task scheduling within parallel systems, to avoid workload imbalances and idle time. 
However, throughput can vary significantly depending on the compression ratio, configuration, and data type, as shown in Figure~\ref{fig:comp-time}.
Failure to accurately estimate compression time can result in performance degradation in real-world applications~\cite{jin2022accelerating}.
Yet predicting compression time is challenging due to the complexity introduced by various factors, including input data characteristics, user configurations, and the runtime hardware environment.
Additionally, the computational proportion of different components in the compression process also varies, as shown in Figure~\ref{fig:comp-time}.
Some studies use empirical methods to estimate compression time, but these approaches often struggle with poor prediction accuracy, scalability, and generalizability~\cite{jin2022improving, jin2024concealing}

To address this issue, we propose a novel analytical method that efficiently and accurately models the time of scientific lossy compression. 
Specifically, we introduce four key strategies:
\textbf{Decompose compression stages} to effectively divides scientific lossy compression into four time-consuming stages: prediction and quantization, frequency counting and codebook construction, Huffman encoding, and other lossless encoding.
\textbf{Theritical analysis} to pinpoint the specific factors contributing to compression time variation by examining computational complexity and establishing models that account for algorithms, compilers, and hardware interactions.
\textbf{Surregate model} to efficiently captures the computational efforts of each compression stage by approximating the behavior of more complex compression algorithms while minimizing overhead.
\textbf{Uncertainty estimation} to systematically analyzes and estimates prediction uncertainty, addressing both algorithmic uncertainty and system-level uncertainty.
Our final product enables accurate prediction of compression time for scientific datasets based on given configurations, while also providing a 95\% confidence interval. 
We evaluated our approach on six real-world scientific datasets with 21 data fields, achieving a prediction error of $\sim$5\% with low prediction uncertainty.

\section{Background}
\label{sec:background}
\subsection{Lossy Compression For Scientific Applications}

Compression is a widely used technique in various systems and frameworks for reducing data sizes and enhancing performance~\cite{cappello2025multifacets}.
Compared to lossless compression, lossy compression can compress data with extremely high compression ratios by losing non-critical information in the reconstructed scientific data.
In recent years, a new generation of lossy compressors for scientific data has been proposed and developed for scientific applications, such as SZ~\cite{di2016fast, tao2017significantly, sz18} and ZFP~\cite{zfp}. These lossy compressors provide parameters that allow users to control the loss of information due to lossy compression precisely. 
Unlike traditional lossy compressors such as JPEG~\cite{wallace1992jpeg}, 
SZ and ZFP are designed to compress floating-point data and can provide a strict error-controlling scheme based on user's requirements.
Generally, lossy compressors provide multiple compression modes, such as the error-bounding mode. 
The error-bounding mode requires users to set an error type, such as point-wise absolute error bound, and a bound value (e.g., $10^{-3}$). The compressor ensures that differences between original and reconstructed data do not exceed the error bound.

Prediction-based lossy compression, such as SZ, have been proven effective for many scientific applications~\cite{cappello2025multifacets}. It involves three main steps:
(1) Each data point's value is predicted based on its neighboring points, using an adaptive, best-fit prediction method.
(2) The difference between the actual value and the predicted value is quantized, based on the user-set compression mode and error bound.
(3) Customized Huffman coding and additional lossless compression are applied to achieve a high compression ratio.

\subsection{Performance Modeling}
Performance modeling involves creating models to predict system behavior under different conditions, helping improve efficiency, scalability, and resource use~\cite{perfmodel}. In high-performance computing (HPC) and large-scale simulations, it is essential for optimizing performance, reducing costs, and streamlining workflows. Understanding the relationship between input parameters and metrics like execution time and memory usage enables better system utilization at scale~\cite{jin2022accelerating}.

While improving performance, compression introduces additional variability into scientific applications due to the uncertainties in execution times and compression ratios.
Although previous studies have proposed effective solutions for predicting compression ratios~\cite{jin2022improving}, predicting compression time remains challenging, with no practical solutions beyond empirical modeling, which highly limits both accuracy and generalizability.
Traditional performance models such as machine learning and regression analysis are impractical for estimating system behavior in scientific lossy compression due to the high number of influencing factors~\cite{mlperf, regperf}.
Furthermore, the prediction method must maintain low overhead to ensure system efficiency. 
To this end, our goal is to design a fast and accurate prediction solution for scientific lossy compression.
\section{Design Methodology}
\label{sec:design}
Based on our observation in Figure~\ref{fig:comp-time}, we carefully decompose prediction-based lossy compression into four time-consuming stages, and establish prediction models for each.
These stages are applicable to most prediction-based scientific lossy compression methods.
We then theoretically analyze the causes of time variance in each stage and propose corresponding equations and/or surrogate models to accurately predict execution time.
To improve the generalizability of our predictions, we eliminate any dependencies specific to a single application. 
On the contrary, many of the parameters used are machine-specific. 
As a result, we develop an offline tuning mechanism for all four stages to efficiently adapt tunable parameters to a given system or environment using a set of sampled data. 
This enables our solution to make efficient predictions during runtime, regardless of the input data type or compression configurations.

\subsection{Prediction And Quantization}
Commonly used predictors in prediction-based lossy compressors—such as the Lorenzo predictor~\cite{tao2017significantly}, regression predictor~\cite{tao2017significantly}, and interpolation predictor~\cite{zhao2021optimizing}—and the quantization process all have a theoretical complexity of O(n). However, in practice, the runtimes of these two steps can vary by up to 20\% with changes in the error bound. Our analysis reveals that this variation is caused by numerous conditional statements in the quantization process. By leveraging sampling and building upon previous research that predicts quantization bin distributions~\cite{jin2022improving}, we can predict the outcomes of these branches. Based on this, we propose a surrogate model using linear regression, which efficiently and accurately predicts the time costs of these steps.
\subsection{Frequency Counting And Codebook Construction}
\label{subsec:3-2}
Variable-length encoding, such as Huffman encoding, significantly reduces data size after prediction and quantization. The preprocessing steps in Huffman encoding include frequency counting and codebook construction. Frequency counting, which theoretically has an $O(N)$ time complexity, is the step where time variations occur in practice. Smaller error bounds lead to larger quantization bins, causing more cache misses during frequency counting and resulting in time variations of up to 100\%. In contrast, codebook construction time is negligible because the codebook size is much smaller than the data size. Similar quantization bin distributions yield comparable time costs and compression ratios, which are predictable~\cite{jin2022improving}. Since quantization bins often exhibit a clear central distribution, we can treat the frequency distribution as an ordered sequence, allowing us to compute Huffman code lengths without constructing the tree, further reducing computational cost. Observing that the time cost of frequency counting varies smoothly with the compression ratio of Huffman encoding, we use the predicted code distribution to calculate the compression ratio. Then, using a moving average on the time cost sequence obtained from offline tests, we accurately predict the time cost for this step. This approach serves as a surrogate model, significantly saving computational effort.
\subsection{Huffman Encoding}
\label{subsec:3-3}
The complexity of encoding is $O(N)$, as it only encodes one element once. However, when the quantization bin distribution becomes less concentrated, the encoded outputs become longer, leading to increased time costs. This increase, however, is not linear. We proposed a probability-based model to explain this behavior.
\subsubsection{Concatenate codes of different lengths}
\label{subsubsec:3-3-1}
\begin{figure}[]
    \centering
   \includegraphics[width=0.65\linewidth]{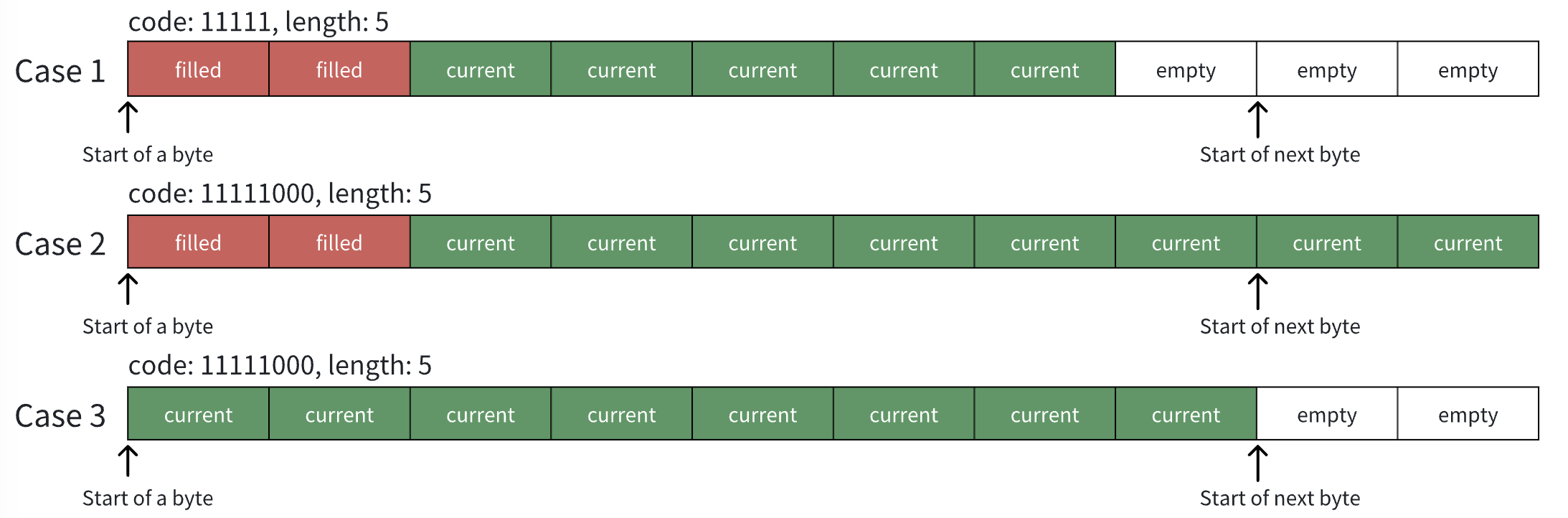}
    \caption{Case 1: to align 1 byte, when the current code is short enough and doesn't need to write new bytes. Case 2: when the current code exceeds the remaining length of the current byte, it will fill the remaining length and then start a new byte. Case 3: the current byte is empty.}
    \label{fig:fig-cases}
\end{figure}

A byte (i.e., 8 bits) is typically the smallest in-memory storage unit.
When concatenate the Huffman codes, operations are performed in bit level, resulting in three cases to accommodate this difference in granularity, as shown in Figure~\ref{fig:fig-cases}.
These cases have different computational cost. As the error bound changes, the proportion of each case shifts, which is the key factor affecting compression time cost difference during this stage.
Since the prediction process effectively removes the structural information of the input data, we assume the prediction errors are independent and identically distributed (i.i.d.). According to the Central Limit Theorem, the sum of i.i.d. variables approaches a normal distribution. 
To determine which case each Huffman code fall into, we must determine the remaining bits in current byte, as shown in Figure~\ref{fig:fig-cases}.
The current bit of this byte is the prefix length of concatenated Huffman codes mod 8.
Given that the data size is relatively large, the resulting normal distribution of prefix length features a significantly wider standard deviation $\sigma$ than 8.
Consequently, the prefix lengths mod 8 will follow a uniform distribution.

To this end, we propose to predict the probabilities of different cases as follows: For case 3, it accounts for 1/8 of the total probability due to the uniform distribution of current bit position in this byte. For case 1, let random variable $X$ represent the code length of the current code and random variable $Y$ represent the result of the current prefix sum mod 8. The probability of this case occurring can be calculated using the following formula: $\sum_{i=0}^{7} \sum_{j=0}^{i} P(X=j) \times P(Y=i)$. The remaining probability will correspond to case 2.

\subsubsection{Impact of black-box behaviors} Since we can predict the proportion of each case accurately, a straightforward prediction method is to use a linear combination of the counts from the three cases to estimate the time cost, and this approach works well when the compression ratio is relatively high. However, as the compression ratio decreases, the prediction accuracy of the time cost become worse. This is due to the numerous conditional statements encountered during the encoding process and the frequent accesses to the codebook. The resulting non-linear changes in time cost, caused by branch mispredictions and cache misses, are difficult to quantify and behave like a black box~\cite{Lin_2019,cacheSurvey}.

Through theoretical analysis and experimental validation, we demonstrate that the observed variations are caused by these three factors. However, separating and quantifying the individual contributions of these factors is complex and introduces variability, making it difficult to consistently obtain reliable results. For this reason, we opt not to use a simulation-based approach. Instead, we use the same design as the surrogate model used in Section~\ref{subsec:3-2} to predict the time cost of encoding.

\subsection{Additional Lossless Encoding}
After Huffman encoding, prediction-based lossy compressor performs an additional lossless compression. We observe that when performing lossless compression on data after Huffman encoding, the compression throughput shows a certain correlation with the compression ratio provided by this lossless compressor. Specifically, when compression ratio is higher, the throughput tends to be higher as well, although not in a linear correlation. As shown in Figure~\ref{fig:stack}, its proportion of the overall time remains small and only increases when the compression ratio is very low. Therefore, we propose to use a piecewise function to fit this correlation. Once we predict the compression ratio of this stage~\cite{jin2022improving} and predict the throughput, we use the predicted output data size from Section~\ref{subsubsec:3-3-1} to predict the time of this stage.
\subsection{Uncertainty Analysis}
\subsubsection{System uncertainty}
In practice, compression time can vary even when configurations and data remain unchanged. This variability arises from both OS-level factors, such as system calls and page faults, and hardware-level uncertainties in cache behavior, including synonym and page crossing issues that disrupt cache consistency and locality\cite{cacheSurvey}. Based on our evaluation, their effect on running time is generally minimal(typically around 2\%).
We identify gamma distribution can be used to fit system uncertainty, as the total running time—considered as the sum of independent random variables from each stage—fits the characteristics of a gamma distribution. However, as we introduced stronger assumptions in the uncertainty analysis of the subsequent algorithm, the normal distribution became a more suitable choice, offering better integration with the model.

\subsubsection{Algorithm uncertainty}
Besides system uncertainty, the prediction model itself also has uncertainty, which we call algorithm uncertainty. At each prediction stage, the error arises from the difference between the predicted and actual values. If we break the actual value into the sum of the actual times for each element and distribute the predicted value equally, the prediction error for each element becomes a constant minus a random variable. Assuming the prediction error is i.i.d., the total error at this stage follows a normal distribution, and the sum of errors across stages, being the sum of normal distributions, is also normally distributed. Thus, we chose to fit algorithm uncertainty using a normal distribution.

\section{Evaluation}
\label{sec:evaluation}

\subsection{Environment Setup}
We conduct our evaluation with SZ3, a modularized prediction-based lossy compressor supporting several predictors. The experiments were conducted in the HPC cluster Owls' Nest at Temple University, each node equipped with four Intel Xeon E5-4655v4 processors featuring 32 physical cores and 1.5 TB of DDR4 memory.
We evaluate our solution on six real-world scientific datasets with 21 data fields.

\subsection{Evaluation of Time Prediction}
Overall, the compression time prediction error is below 10\% for all 21 data fields, with compression ratios ranging from 5$\times$ to 1000$\times$, and an average prediction error of 5\%,
Shown in Table~\ref{table:prediction-error}. 
Note that as mentioned in Section~\ref{sec:design}, we only use sampled data to train the tunable parameters of our surrogate model offline, in this case the temperature field from Nyx and the QS field from SCALE.
Specifically, Table~\ref{table:prediction-error} shows the prediction error when using the Lorenzo predictor and the Interpolation predictor. 
As shown on Figure~\ref{fig:comp-time}, prediction \& quantization stage is the most time-consuming of the four stages, accounting for 60\%-70\% of the entire compression time. 
Our solution accurately predicts the behavior of this stage, resulting in the smallest prediction error among all four stages.
We acknowledge that the prediction error for the lossless compression stage is relatively high. However, since it only contributes to $\sim$5\% of the compression time, the impact of this prediction error is minimized.

\definecolor{blue0}{RGB}{150, 150, 255}
\definecolor{blue1}{RGB}{50, 50, 205}
\definecolor{blue2}{RGB}{0, 0, 105}

\cboff
\begin{table*}[]
\centering
\caption{Compression Time Prediction Error}
\tiny
\label{table:prediction-error}
\begin{tabular}{@{}cccccccc@{}}
\toprule
Name                       & Field               & Dim          & \multicolumn{1}{c}{\begin{tabular}[c]{@{}c@{}}Pred \&\\ Quant Err.\end{tabular}} & \multicolumn{1}{c}{\begin{tabular}[c]{@{}c@{}}Freq \&\\ Build Err.\end{tabular}} & Encode Err. & Lossless Err. & Overall Err. \\ \midrule
\multirow{6}{*}{Nyx}       & dark\_matter\_density & 512x512x512  & 1.81\%            & 11.05\%                & 6.35\%      & 17.95\%       & 2.58\%       \\
                           & baryon\_density      & 512x512x512  & 3.34\%             & 15.98\%                & 9.74\%      & 36.99\%       & 4.78\%       \\
                           & temperature         & 512x512x512  & 1.09\%             & 5.65\%                 & 5.18\%      & 12.71\%       & 2.02\%       \\
                           & velocity\_x         & 512x512x512  & 3.76\%             & 18.75\%                & 9.06\%      & 25.24\%       & 4.08\%       \\
                           & velocity\_y         & 512x512x512  & 2.85\%             & 18.93\%                & 7.53\%      & 21.47\%       & 3.16\%       \\
                           & velocity\_z         & 512x512x512  & 3.76\%             & 16.55\%                & 7.10\%      & 22.40\%       & 3.08\%       \\
\multirow{3}{*}{Scale}     & QR                  & 98x1200x1200 & 1.55\%             & 7.60\%                 & 13.51\%     & 46.24\%       & 1.90\%       \\
                           & QS                  & 98x1200x1200 & 1.97\%             & 9.01\%                 & 15.66\%     & 61.51\%       & 2.94\%       \\
                           & U                   & 98x1200x1200 & 3.74\%             & 4.83\%                 & 8.63\%      & 44.57\%       & 3.67\%       \\
\multirow{5}{*}{CESM(3D)}  & CLOUD               & 26x1800x3600 & 4.48\%             & 17.18\%                & 12.49\%     & 55.58\%       & 4.82\%       \\
                           & T                   & 26x1800x3600 & 4.53\%             & 14.39\%                & 10.02\%     & 50.07\%       & 4.40\%       \\
                           & U                   & 26x1800x3600 & 4.49\%             & 11.31\%                & 9.86\%      & 47.33\%       & 3.83\%       \\
                           & OMEGA               & 26x1800x3600 & 4.19\%             & 11.31\%                & 5.81\%      & 46.94\%       & 4.29\%       \\
                           & DTV                 & 26x1800x3600 & 2.88\%             & 9.31\%                 & 7.92\%      & 53.39\%       & 4.03\%       \\
\multirow{2}{*}{HACC}      & yy                  & 1073726487   & 1.21\%             & 16.31\%                & 23.87\%     & 36.70\%       & 5.30\%       \\
                           & xx                  & 1073726487   & 1.23\%             & 16.80\%                & 23.42\%     & 37.57\%       & 5.08\%       \\
\multirow{3}{*}{Hurricane} & TCf                 & 100x500x500  & 3.86\%             & 11.30\%                & 2.53\%      & 11.56\%       & 7.59\%       \\
                           & CLOUDf              & 100x500x500  & 2.01\%             & 4.89\%                 & 10.68\%     & 20.86\%       & 7.65\%       \\
                           & QRAINf              & 100x500x500  & 2.49\%             & 6.17\%                 & 9.53\%      & 21.81\%       & 4.99\%       \\
\multirow{2}{*}{CESM(2D)}  & TSMX                & 1800x3600    & 4.99\%             & 9.74\%                 & 9.36\%      & 72.42\%       & 9.78\%       \\
                           & CLDHGH              & 1800x3600    & 6.02\%             & 14.55\%                & 11.46\%     & 52.92\%       & 8.55\%       \\ \midrule
\end{tabular}
\end{table*}
\cbon


\begin{figure}[]
    \centering
    \begin{subfigure}{0.47\linewidth}
        \includegraphics[width=\linewidth]{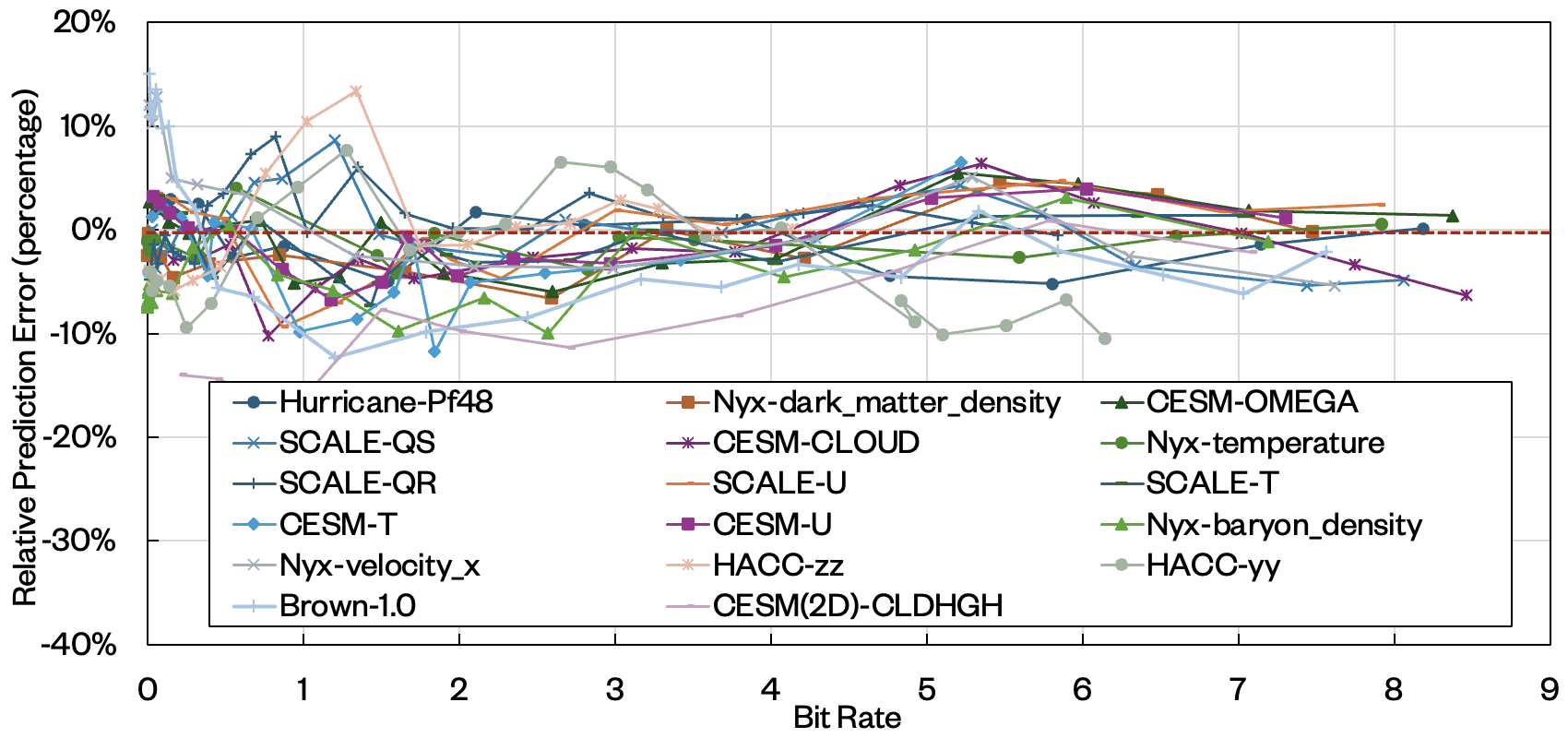}
        \caption{\scriptsize Lorenzo}
        \label{fig:lorenzo-error-overview}
    \end{subfigure}
    \begin{subfigure}{0.47\linewidth}
        \includegraphics[width=\linewidth]{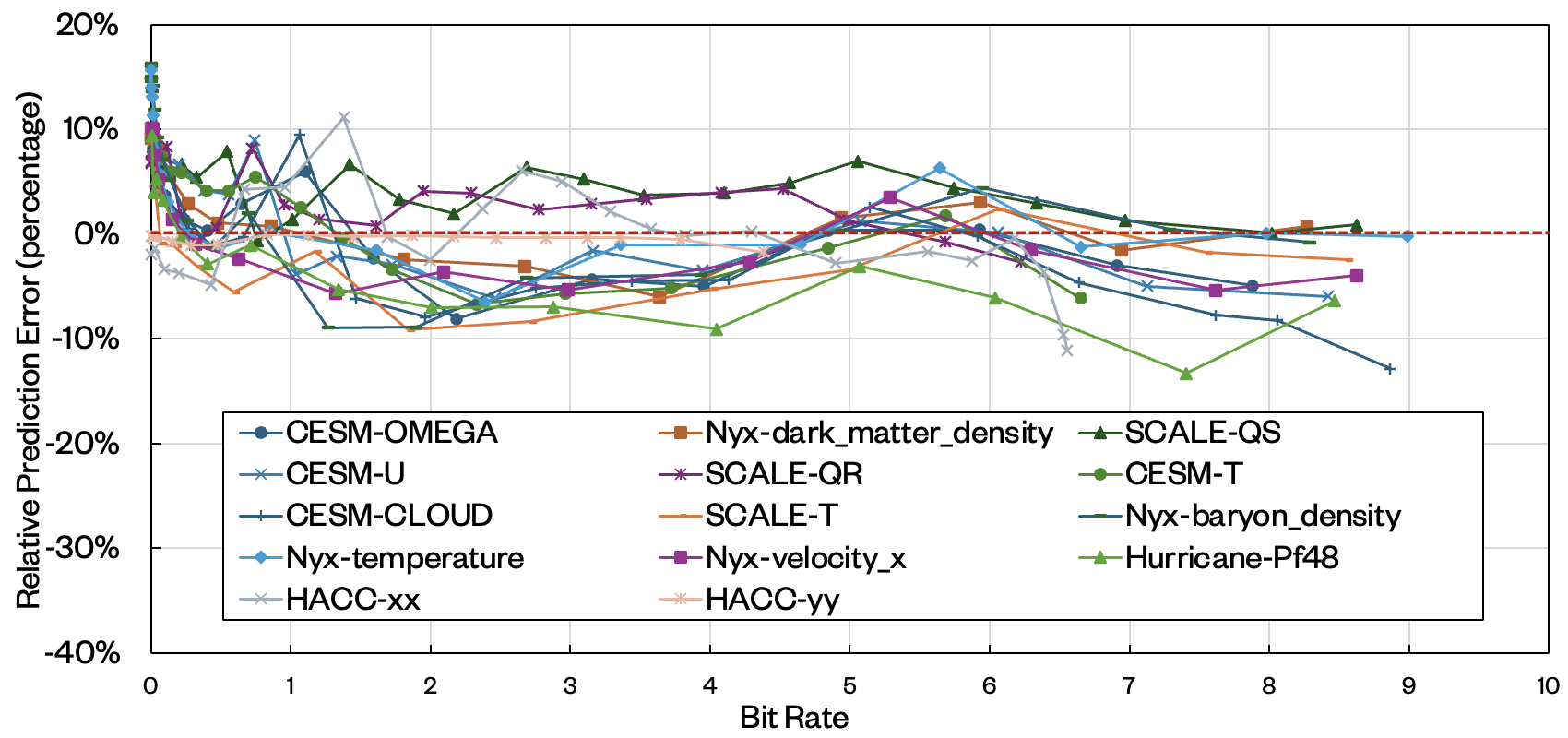}
        \caption{\scriptsize Interpolation}
        \label{fig:interp-error-overview}
    \end{subfigure}
    \caption{Overview of the prediction error for various fields in different datasets}
    \label{fig:error-overview}
\end{figure}

Next, we conduct comprehensive evaluations across various compression ratios using the Nyx dataset contains six data fields.
As shown in Figure~\ref{fig:overall-scale}, our solution accurately predicts compression time across different compression ratios.
Note that the far left and right sides of the figure reflect rarely used compression configurations with extremely low or high compression ratios, demonstrating that our solution remains robust in extreme scenarios.


\begin{minipage}{0.4\linewidth}
    \centering
    \vspace{-3mm}
    \includegraphics[width=\linewidth]{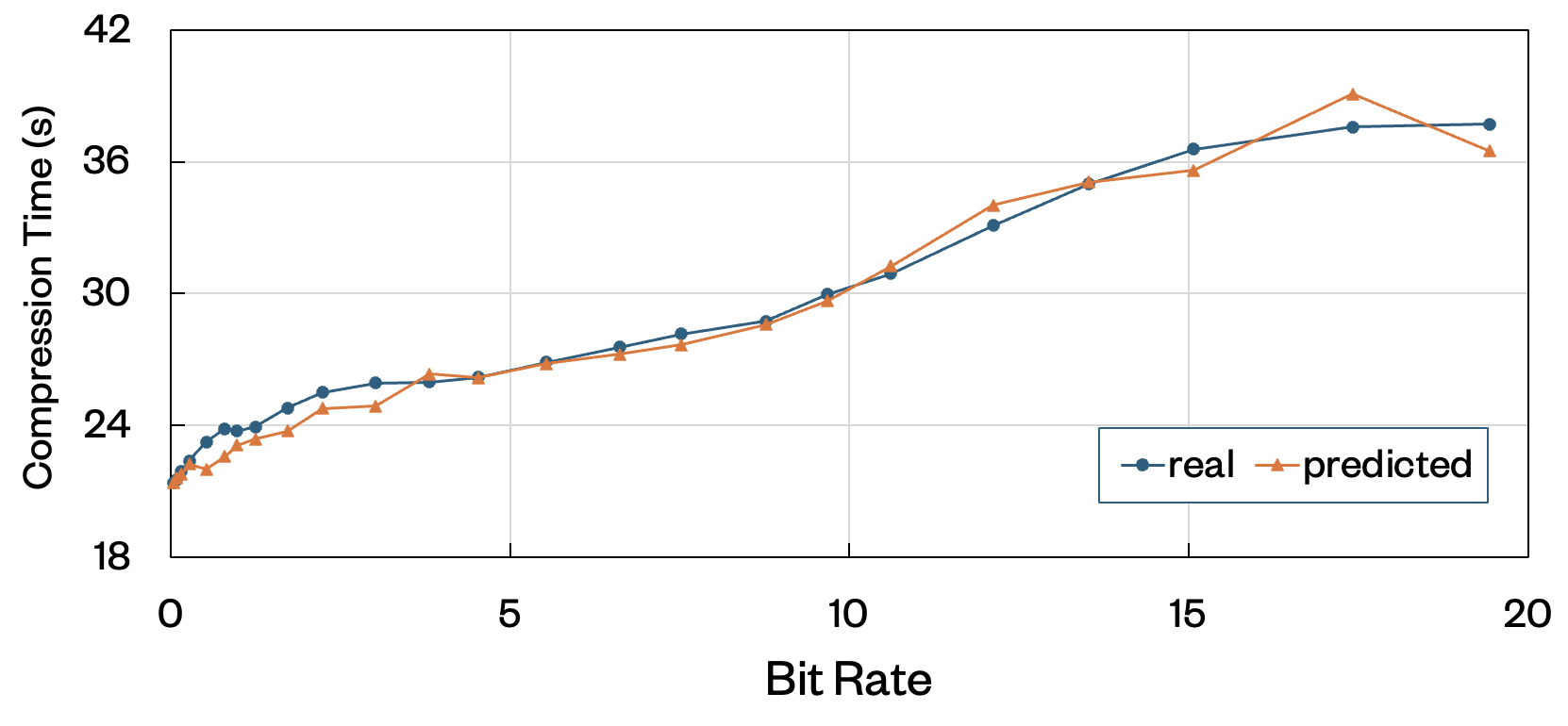}
    \vspace{-7mm}
    \captionof{figure}{Comparison between real and predicted time across bitrate.}
    \label{fig:overall-scale}
\end{minipage}%
\hspace{1.5em}
\begin{minipage}{0.4\linewidth}
    \centering
    \tiny
    \begin{tabular}{@{}cccc@{}}
        \toprule
        Name      & 95\% CI lower bound & 95\% CI upper bound & $\sigma$ \\ \midrule
        Nyx       & -8.48\%             & 6.82\%              & 3.9\% \\
        SCALE     & -7.77\%             & 7.48\%              & 3.9\% \\
        CESM(3D)  & -10.56\%            & 6.72\%              & 4.4\% \\
        CESM(2D)  & -22.88\%            & 24.87\%             & 4.5\% \\
        Hurricane & -13.01\%            & 5.69\%              & 4.7\% \\
        HACC      & -13.11\%            & 5.60\%              & 4.8\% \\ \midrule
    \end{tabular}
    \vspace{0.4em}
    \captionof{table}{$95\%$ Confidence interval for different datasets.}
    \label{table:ci}
\end{minipage}
\vspace{-3mm}


\subsection{Evaluation of Uncertainty}
\begin{figure}[h]
    \centering
    \begin{subfigure}{0.24\textwidth}
        \centering
        \includegraphics[width=\textwidth]{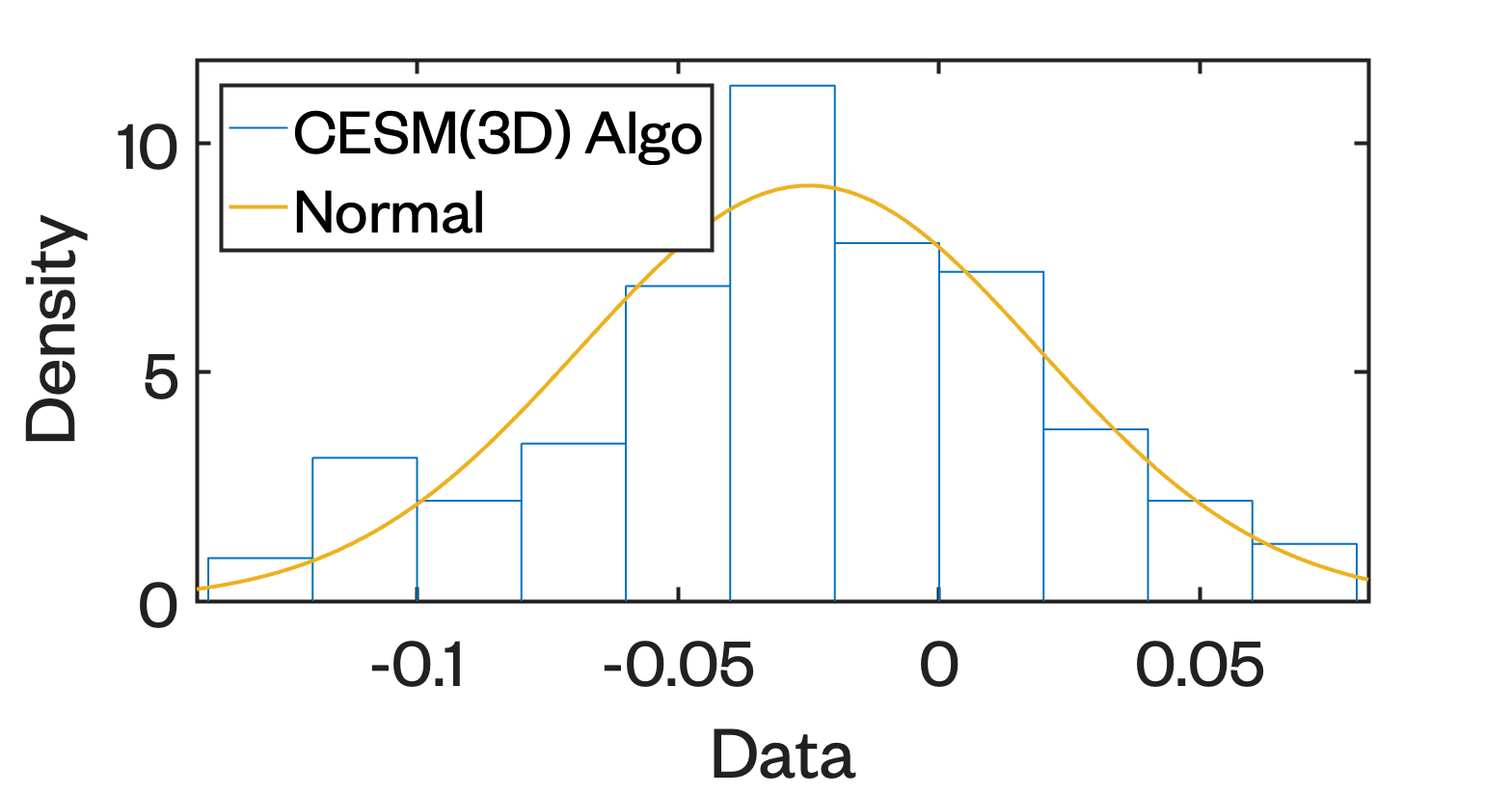}
        \caption{\scriptsize CESM(3D) Algo\_Uncer}\label{subfig:cesm-algo}
    \end{subfigure}
    \begin{subfigure}{0.24\textwidth}
        \centering 
        \includegraphics[width=\textwidth]{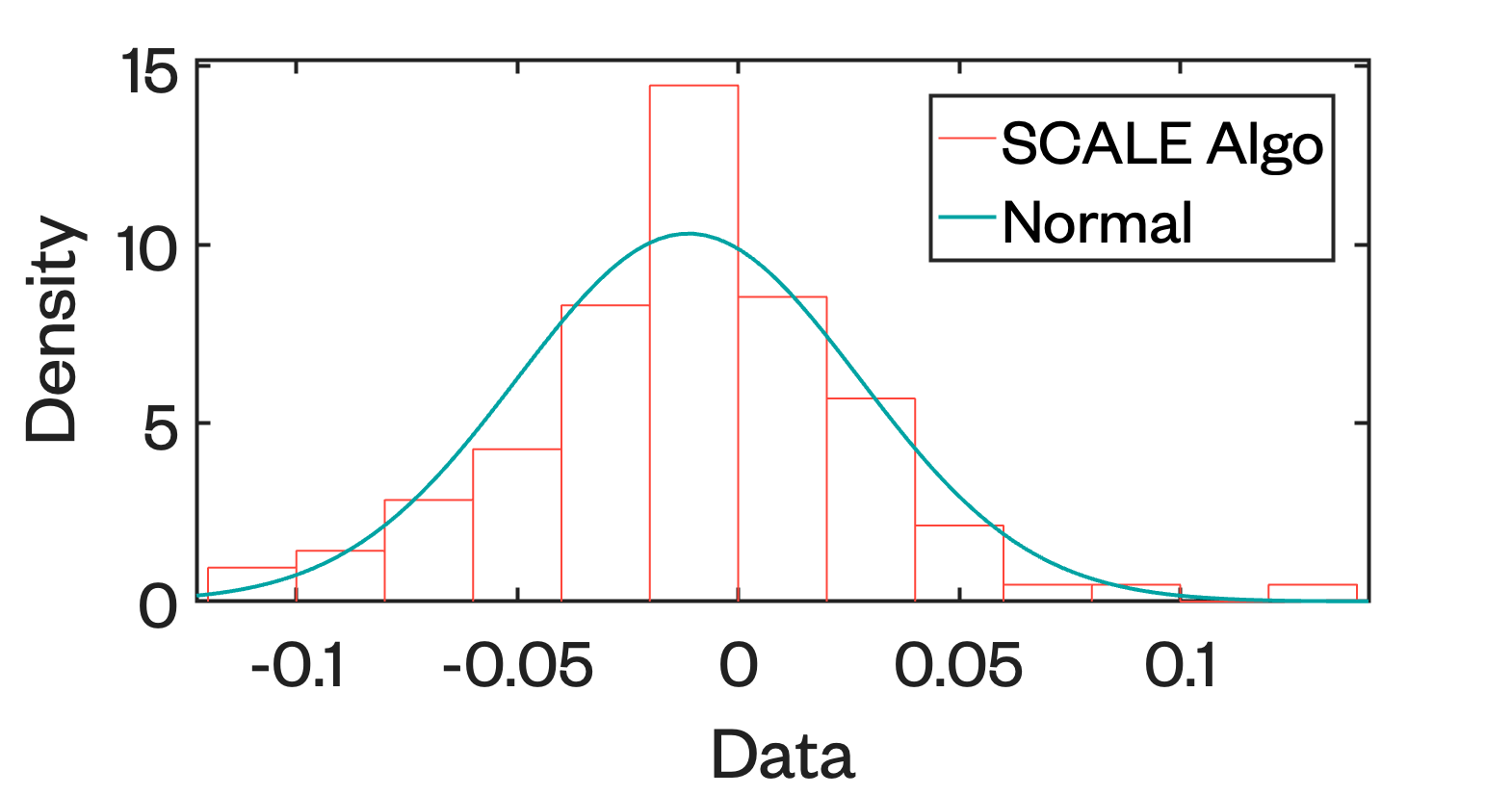}
        \caption{\scriptsize SCALE Algo\_Uncer}\label{subfig:scale-algo}
    \end{subfigure}
    \begin{subfigure}{0.24\textwidth}
        \centering
        \includegraphics[width=\textwidth]{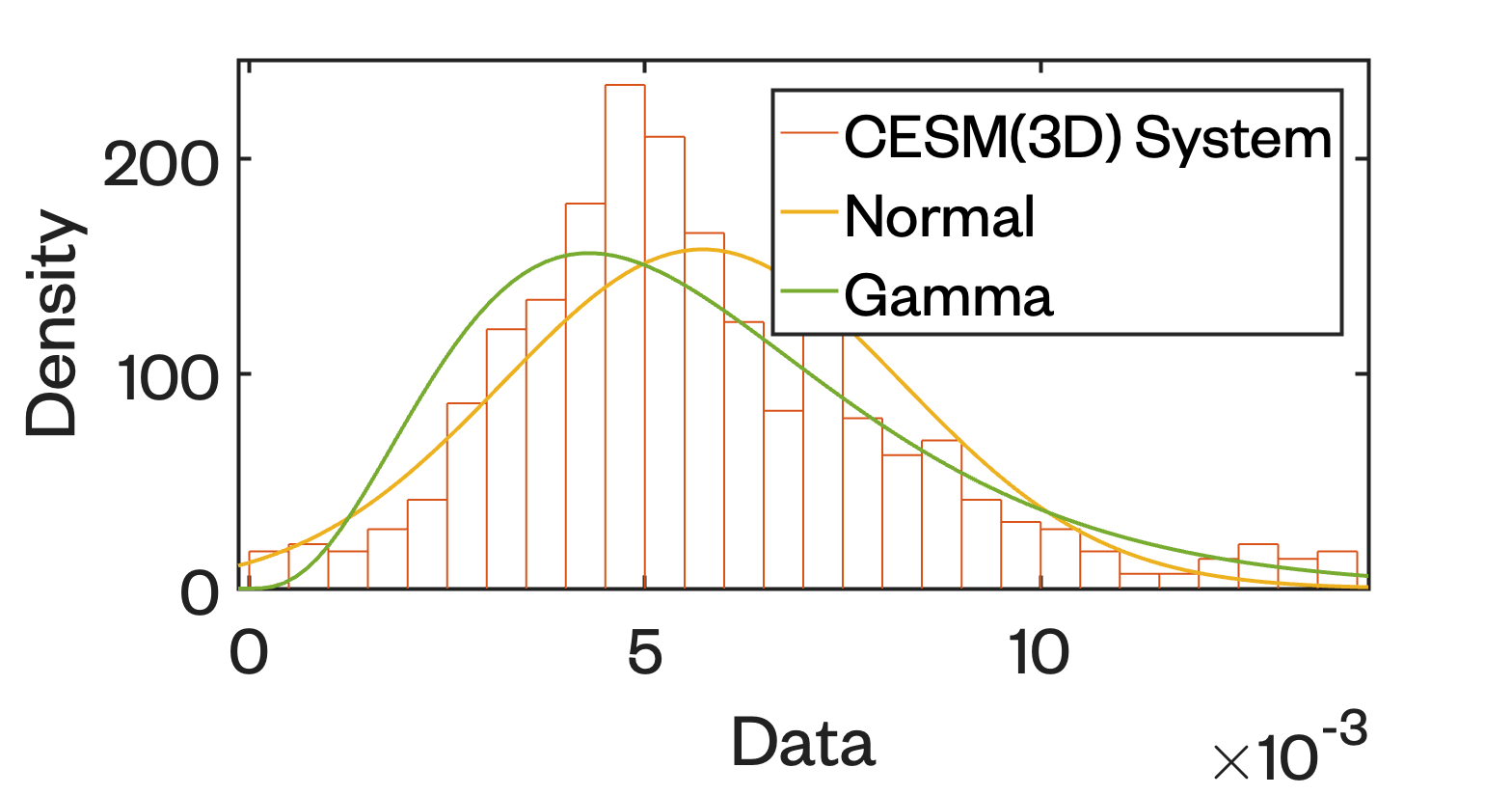}
        \caption{\scriptsize CESM(3D) Sys\_Uncer}\label{subfig:cesm-sys}
    \end{subfigure}
    \begin{subfigure}{0.24\textwidth}
        \centering 
        \includegraphics[width=\textwidth]{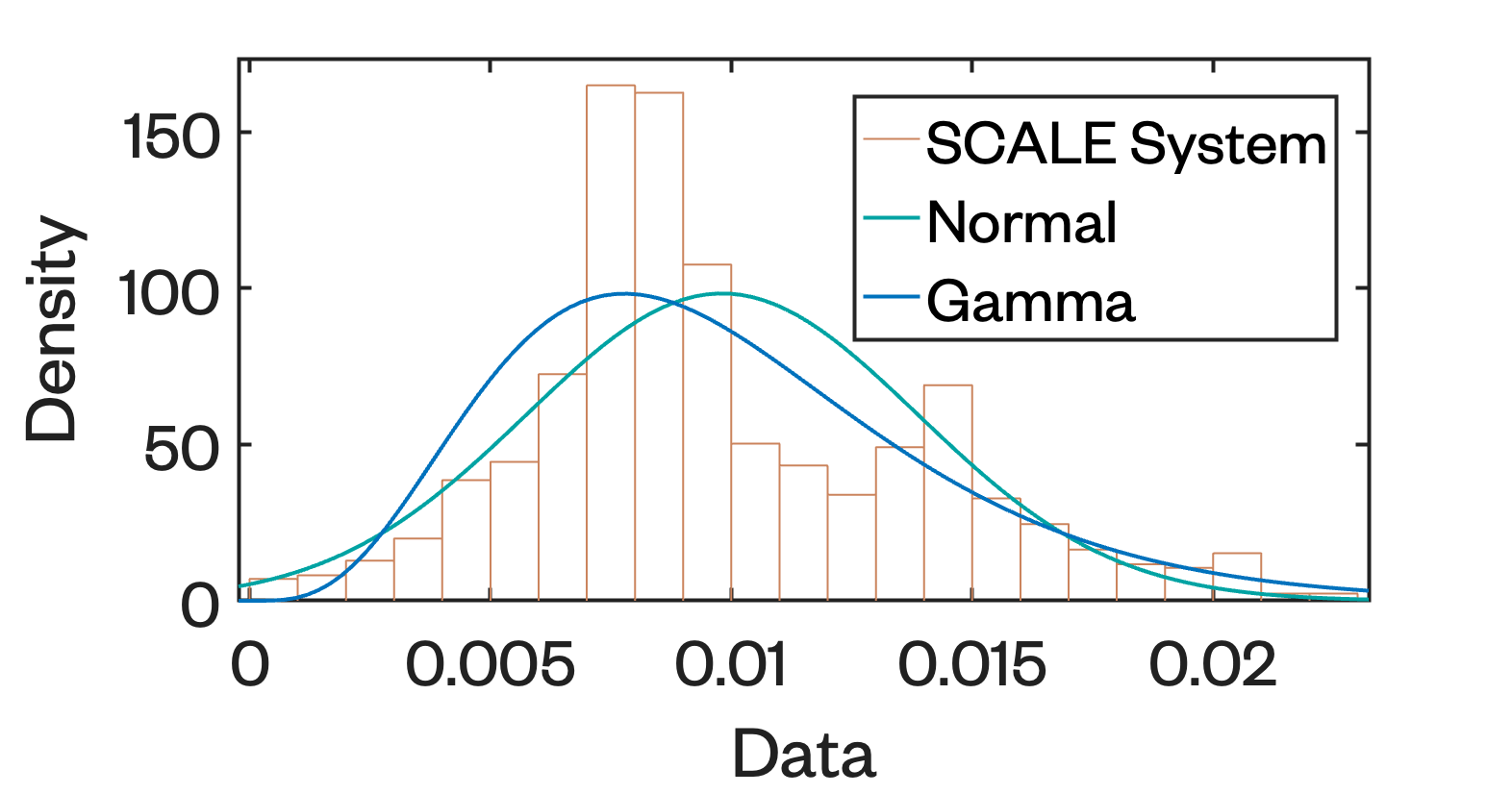}
        \caption{\scriptsize SCALE Sys\_Uncer}\label{subfig:scale-sys}
    \end{subfigure}
    \caption{The algorithm and system uncertainty distribution and its corresponding gamma distribution and normal distribution fit for CESM dataset and SCALE dataset.}
    \label{fig:uncertainty}
\end{figure}

As mentioned in section 3.5.2, we propose modeling both algorithmic and system uncertainties using a normal distribution.
Across all datasets, both uncertainties exhibit a clear central distribution pattern, with the standard deviations $\sigma$ of the fitted normal distributions being of the same order of magnitude. 
Figure~\ref{fig:uncertainty} shows the actual uncertainty distribution alongside the fitting results for both the CESM (3D) and SCALE datasets, demonstrating a strong alignment between the observed data and the fitted normal distribution.
\vspace{-1em}
\vspace{-1em}
As both algorithmic and system uncertainties are modeled using normal distributions, the total uncertainty can be directly obtained by summing the two distributions. Therefore, the final confidence interval can also be directly calculated from the combined uncertainties. The $95\%$ confidence intervals for the six datasets used in the experiments are shown in Table~\ref{table:ci}.
\subsection{Case Study}
\subsubsection{Predictor Selection}
Optimal compression performance, in terms of balancing compression ratio, throughput, and data quality, can be achieved by selecting the most appropriate predictor based on the given conditions. 
Taking the U field in CESM as an example. When the error bound is between $1 \times 10^{-2}$ and $5 \times 10^{-5}$, interpolation is the preferred solution with high throughput and high compression ratio. Whereas between $5 \times 10^{-5}$ and $2 \times 10^{-6}$, Lorenzo is slightly slower than interpolation but has a higher compression ratio. When drops below $2 \times 10^{-6}$, Lorenzo is the preferred solution with high throughput and high compression ratio.
Our solution enables the estimation of these outcomes without executing the compression to benchmark, thereby allows adaptive selection of the optimal predictor.

\subsubsection{Error Bound Selection}
In most compression scenarios,
compression time tends to increase as the compression ratio decreases. 
Using the binary search method and the proposed prediction model, we are able to efficiently identify the maximum applicable error bound for a target compression time with minimal overhead.

\subsection{Prediction Performance}

The complexity of the compression algorithm is $O(N + K \log K)$, where $N$ is the data size and $K$ is the quantization bin size. The complexity of running the prediction model is $O(NR)$, where $R$ is the sampling rate. 
The prediction model skips the steps of Huffman encoding, building the Huffman tree, and lossless compression. 
Although it includes an additional sampling step, it does not have a significant negative impact on the overall runtime. Based on our evaluation, the time required to perform the prediction is almost negligible. 
For example, in the baryon\_density field of the $512\times512\times512$ Nyx dataset, the time taken for sampling and prediction is on average 3.9\% of the total compression time, with a sampling rate of 4\%. 
Across all datasets evaluated, the average sampling and prediction cost is approximately 4.5\% of the total compression time, with a consistent sampling rate of 4\%. 

\section{Conclusion and Future Work}
\label{sec:conclusion}
In this paper, we propose a general and accurate solution for predicting compression time of scientific lossy compression. 
We decompose the compression process into four stages, then theoretically analyze the causes of time variance and design equations and/or surrogate models to eliminate data dependencies.
Our experiments demonstrate that our solution accurately predicts compression time across six real-world datasets, with a prediction error of $\sim$5\%. We also provide a 95\% confidence interval for these predictions. 
In the future, we plan to refine the model to support a broader range of compression algorithms. We also aim to integrate this solution into real-world scientific applications to enable compression time-aware task scheduling and resource allocation.


\section*{Acknowledgments}

{
\footnotesize 
This research includes calculations carried out on HPC resources supported in part by the National Science Foundation through major research instrumentation grant number 1625061 and by the US Army Research Laboratory under contract number W911NF-16-2-0189. The data resources is provided on SDRBench~\cite{sdrbench,Zhao_2020}.
}
\bibliographystyle{IEEEbib}
\bibliography{refs}










\end{document}